\documentclass[prl,twocolumn,showpacs]{revtex4}

\usepackage{graphicx}
\usepackage{amsmath}

\newcommand{\kbar}{\mathchar'26\mkern-9mu k}

\begin{document}
\title{Experimental Realization of a Quantum Ratchet Through Phase Modulation}
\author{D.~H.~White, S.~K.~Ruddell and M.~D.~Hoogerland\footnote{E--mail: m.hoogerland@auckland.ac.nz}}
\affiliation{Department of Physics, University of Auckland, Private Bag 92019, Auckland, New Zealand}

\begin{abstract}
We report on an experimental realization of unidirectional transporting island structures in an otherwise chaotic phase space of the delta--kicked rotor system. Using a Bose-Einstein Condensate as a source of ultracold atoms, we employ asymmetric phase modulation in the kicks, with the narrow momentum distribution of the atoms allowing us to address individual island structures. We observe quantum ratchet behavior in this system, with clear directed momentum current in the absence of a directional force, which we characterize and connect to $\epsilon$--classical theory. 
\end{abstract}
\maketitle

The concept of a Brownian ratchet, which allows Brownian motion to do work, has been considered as a thought experiment by von Smoluchowski~\cite{Smol1912}, and analyzed by Feynman~\cite{Feyn1962}. There has been reinvigorated interest in such ratchets in recent years due to potential biological applications~\cite{Matthias2003,Chowdhury2005}. The great progress made in the control of ultracold atoms~\cite{Bloch2008} has allowed for a new viewpoint into ratchet systems~\cite{Dana2008}. 

Brownian motion is closely related to microscopic chaos~\cite{Gaspard1998}. While there is some debate as to whether or not Brownian motion is a direct result of underlying chaos~\cite{Cecconi2005}, the similarities between the random--walk type evolution exhibited by chaotic trajectories and Brownian particles are numerous~\cite{Kaplan1993}. The field of quantum chaos examines such microscopic chaotic systems in the quantum limit~\cite{Izrailev1990}. The delta--kicked rotor has become established as a workhorse in this area, with an atom optics implementation pioneered by Raizen \emph{et. al.}~\cite{Moore1995,Raizen1999}. Although a conceptually simple system, it is capable of exhibiting rich and complex dynamics. Capitalizing on the narrow momentum distribution of a Bose--Einstein condensate (BEC) provides a clean method of examining a quantum chaotic system and allows for precision experiments analyzing the transition from quantum to classical behaviors~\cite{Ullah2011}.

Directed transport in an unbiased potential (that is, a potential periodic in space and time) must be achieved through broken symmetry. Previous investigations have primarily focused on modified initial conditions~\cite{Dana2008}, and on potentials asymmetric in space~\cite{Salger2009} and in time~\cite{Wang2008b}. Ratchets with classical analogs require mixed phase spaces: regular, integrable regions embedded in a chaotic sea~\cite{Schanz2001}. A number of other interesting systems possess mixed phase spaces. The accelerator modes of the kicked rotor, manifesting in the case of an accelerated lattice, are an example of a mixed phase space that has been experimentally studied~\cite{Oberthaler1999,Behinaein2006}. Linearly accelerated atoms result from classical trajectories launched from the accelerator mode phase space islands.

The experiments that we report on in this Letter have been motivated by recent interest in a phase modulated kicked rotor system. It has been shown that islands of stability similar to the accelerator modes can be induced through a phase modulation~\cite{Gong2003}. Gong \emph{et al.} suggested reversing the sign of the kicking potential every two kicks or, equivalently, utilizing a phase modulation set of \{0,0,$\pi$,$\pi$\}. Theoretical studies found this to produce paired islands, accelerating atoms in opposite directions. Recent experiments utilizing cold thermal atoms have shown broadening of the momentum distribution, indicating enhanced transport, when a phase modulation set of \{0,$\frac{\pi}{2}$,0,$-\frac{\pi}{2}$\} was utilized~\cite{Sadgrove2013}.

We experimentally investigate a phase--modulated $\delta$--kicked rotor system that has a mixed pseudo--classical phase space, with transporting islands of stability similar to those of the accelerator modes. There is, however, an important difference in our system. The accelerator modes exist only in the presence of a net biased force (e.g. gravity). Our system consists only of a pulsed lattice with a periodic phase modulation, for which there is no net force. We propose an asymmetric phase modulation set of \{0,$\frac{2\pi}{3}$,0\} to induce broken symmetry and instigate ratchetlike directed transport, despite the apparent directional neutrality. Use of a Bose--Einstein Condensate (BEC) enables us to explore the phase space and probe individual islands of stability. 

We employ the atom--optics kicked rotor (AOKR), which consists of atoms subjected to a pulsed standing wave created by off--resonant laser light. Consider a generalized AOKR model with $N$ kicks, in dimensionless units, with scaled time--dependent Hamiltonian:

\begin{equation}
H(t^{\prime}) = \frac{\kbar\rho^2}{2} + k\sum_{i=0}^{N-1} \cos(\vartheta+\phi_i) \delta(t^{\prime}-i).
\label{eq:hamiltonian}
\end{equation}

\noindent The absolute phase of the standing wave at the $j^{th}$ kick is given  by $\phi_j = \varphi_{j~\text{mod}~n}$, where we utilize the phase modulation set $\{\varphi_0,\varphi_1,\ldots,\varphi_{n-1}\}$. We utilize the following scaled units: $\vartheta = 2k_Lx$; $\rho = p/2p_r$; $\kbar = 4\hbar k_L^2T/m$; $k = P\tau\Gamma^2/(8\pi|\delta| I_{sat}r^2)$; and $t^{\prime} = t/T$. Here, $k_L$ is the wavenumber of the kicking laser; $p_r = \hbar k_L$ is one photon recoil unit of momentum; $m$ is the atomic mass; $T$ is the pulse period; $\tau$ is the duration of the pulse; $P$ is the power in each beam; $\Gamma$ is the natural linewidth; $\delta$ is the laser detuning from resonance; $r$ is the laser beam--waist radius; and $I_{sat}$ is the saturation intensity. 

We write $\kbar = 2\pi\ell + \epsilon$, where integer $\ell$ determines the resonance order and $|\epsilon| \ll \pi$.  For $^{87}$Rb, $\kbar = 2\pi$ corresponds to a pulse period of 33.1 $\mu$s. With 50 mW of power in each beam, a beam--waist of 110 $\mu$m and a detuning of 100 GHz, a pulse duration of 300 ns produces a kick--strength of $k = 4.5$.

In the vicinity of a quantum resonance of the kicked rotor, the dynamics can be described by a fictitious classical limit~\cite{Fishman2002}, known as $\epsilon$--classical theory. If $|\epsilon|$ is regarded as Planck's constant, it can be shown that the time evolution operator for the system described by (\ref{eq:hamiltonian}) is the formal quantization of the classical standard map~\cite{Wimberger2004,Sadgrove2013,Sadgrove2011}:

\begin{align}
	&J_{i+1} = J_i + k|\epsilon|\sin(\theta+\varphi_i), \\
	&\theta_{i+1} = \theta_i + J_{i+1},
\end{align}

\noindent where the dimensionless generalized momentum $J = \epsilon\rho + \pi\ell + 2\pi\ell\beta$ and $\theta = \vartheta +\pi[1-\text{sgn}(\epsilon)]/2$. The fractional part of the momentum, also known as the quasi--momentum, is $\beta$. The map is dependent only upon the parameters $k|\epsilon|$ and the phases $\varphi_i$.

\begin{figure}
\includegraphics[width=250px]{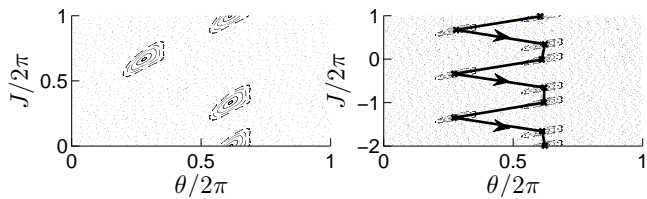}
\caption{Poincar\'{e} sections of transporting islands for phase modulation set $\{0,\frac{2\pi}{3},0\}$, $k|\epsilon| = 3.2$. Islands embedded in the chaotic sea are shown in (a). A visualization of the transporting nature of these islands is provided in (b), showing an example classical trajectory launched from a transporting island [with initial co--ordinates $(\theta,J) = (3.8, 6.2)$].}
\label{fig1}
\end{figure}

In Fig.~\ref{fig1}(a) the Poincar\'{e} section for the $\{0,\frac{2\pi}{3},0\}$ phase set is drawn. One notable feature is that the attractors, or islands, visible in the phase portrait are much larger than those previously investigated~\cite{Gong2003,Sadgrove2013}. Another particularly interesting feature is that these islands are unidirectional in momentum, as indicated in Fig.~\ref{fig1}(b), unlike those from the $\{0,0,\pi,\pi\}$ set which come in pairs of opposite directions~\cite{Gong2003}. This unidirectionality is induced by the asymmetry of the phase modulation set and immediately suggests a new ratchet mechanism.

It should be noted that although all three of the islands in Fig.~\ref{fig1}(a) are involved in transportation, for our chosen phase modulation transportation only occurs if the trajectory is launched from an island centered around integer multiples of $J = 2\pi$. Different islands can become transporting launching pads by using a circular permutation of the phase modulation set. To access an island centered at $J_0$, a quasi--momentum of $\beta = (J_0-\pi\ell)/(2\pi\ell+\epsilon)$ is required. For $\epsilon = \ell = 1$ and $J_0 = 2\pi$ we require $\beta = 0.43$.

\begin{figure}
\includegraphics[width=250px]{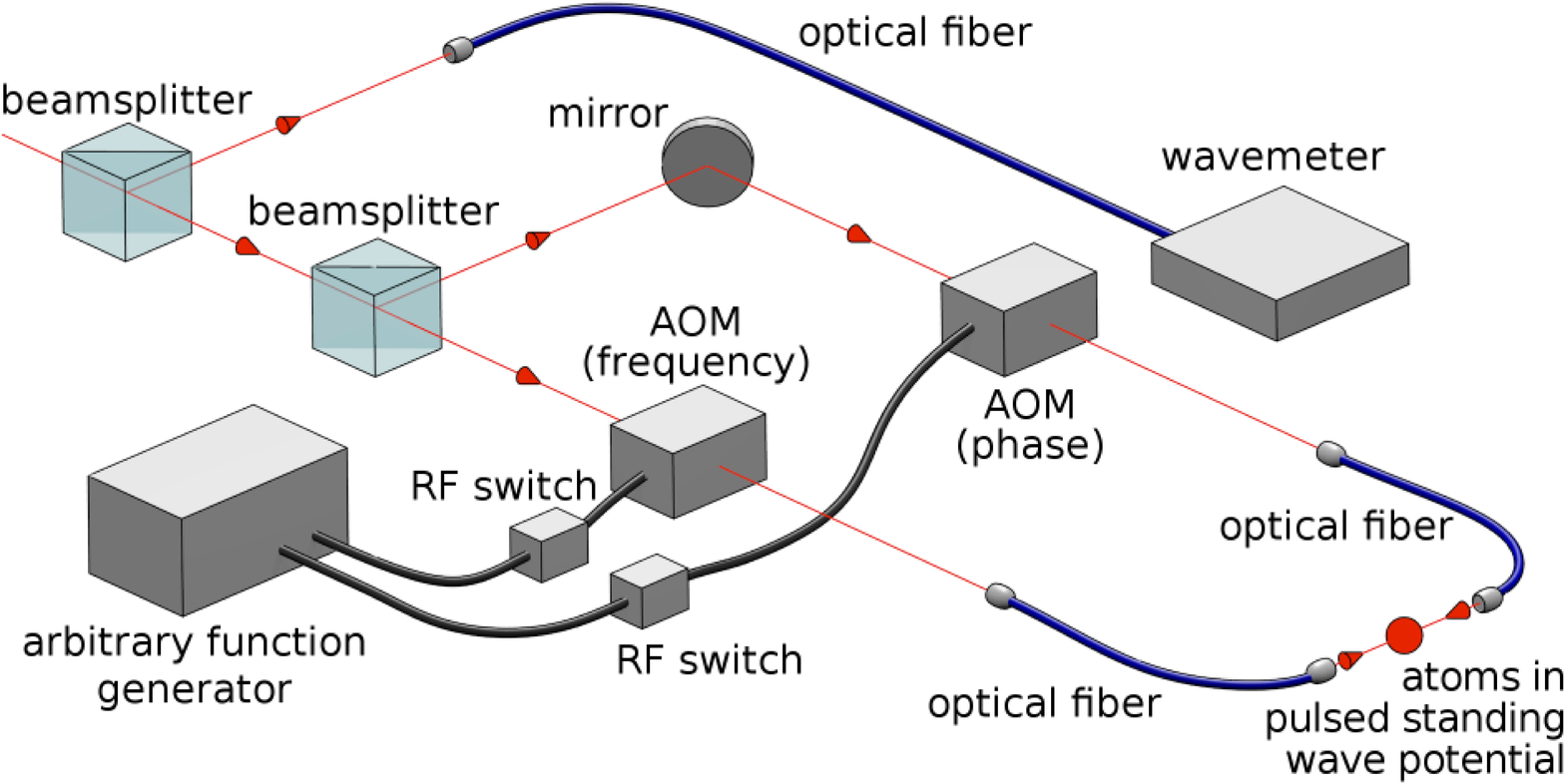}
\caption{(Color online) Experimental scheme. A $^{87}$Rb Bose--Einstein condensate is subject to a phase--modulated pulsed standing wave. The standing wave is created by two counterpropagating linearly polarized laser beams detuned by 100 GHz from resonance. The left channel of the arbitrary function generator (AFG) controls an acousto--optic modulator (AOM) to modulate the phase of the standing wave; the right channel controls the frequency difference between the beams, thereby creating an initial condition of non--zero atomic momentum. A two--channel programmable pulse generator (not shown) controls both of the RF switching amplifiers that toggle the AOMs, in addition to the phase modulation input to the AFG. The wavemeter allows us to monitor the detuning of the laser from resonance.}
\label{fig2}
\end{figure}

Our experimental setup is shown in Fig.~\ref{fig2}. The starting point for the experiment is a $^{87}$Rb BEC of $\approx$ 30000 atoms, prepared in the $|F = 1\rangle$ hyperfine state by evaporative cooling in an all--optical BEC apparatus~\cite{Wenas2008}. The BEC is allowed to freely expand for 1.0 ms before kicking to minimize nonlinear effects during the kicking sequence. A pulsed optical standing wave is then applied via two linearly polarized counterpropagating laser beams detuned by 100 GHz to the red from the $5S_{1/2} F = 1 \rightarrow 5P_{3/2} F = 2$ transition. The large detuning minimizes decoherence effects associated with spontaneous emission. The detuned optical standing wave results in the sinusoidal term in the Hamiltonian (\ref{eq:hamiltonian}) due to the AC Stark shift. The amplitude, phase and frequencies of the two beams are independently controlled via acousto--optic modulators (AOMs), driven by a Tektronix Arbitrary Function Generator (AFG3252). Following the kick sequence, the atoms are allowed to expand for 3 ms and are then imaged using absorption imaging.

The AFG3252 is able to arbitrarily modulate the standing wave phase with a response time of 35 $\mu$s, allowing us to consider a positive--$\epsilon$ situation around the $\ell = 1$ resonance, or any $\epsilon$ for higher order resonances. A two--channel pulse generator signals RF switches to turn the AOMs on and off, and provides an analog signal to the phase modulation input of the AFG3252. A small frequency shift $\delta_{\omega}$ is applied to one beam to create a moving standing wave in the laboratory frame; in the frame of the standing wave, the atoms of mass $m$ have an initial momentum (measured in two--photon recoil units) given by $\beta = m\delta_{\omega}/(4\hbar k_L^2)$~\cite{Currivan2009}. $\beta = 0.43$ corresponds to a 12.9 kHz frequency difference.

Utilization of a BEC provides a very narrow initial spread of momenta. After mean--field repulsion is considered, the BEC has an initial spread in momentum space of 0.5 $\hbar k_L$, or 25\% of a phase space cell. The BEC is completely delocalized in $\theta$. We are therefore able to individually address islands in phase space separated in $J$ by controlling the initial momentum $\beta$.

In order to compare our experimental data with theory, we conduct Schr\"{o}dinger equation simulations via the split--step method, with the Hamiltonian in (\ref{eq:hamiltonian}). The simulations are conducted in one dimension. We use an initial condition of a Gaussian state of 30000 $^{87}$Rb atoms with initial position spread of 1.6~$\mu$m that has undergone mean--field repulsion for 1.0 ms, as in the experiment.

\begin{figure}
\includegraphics[width=250px]{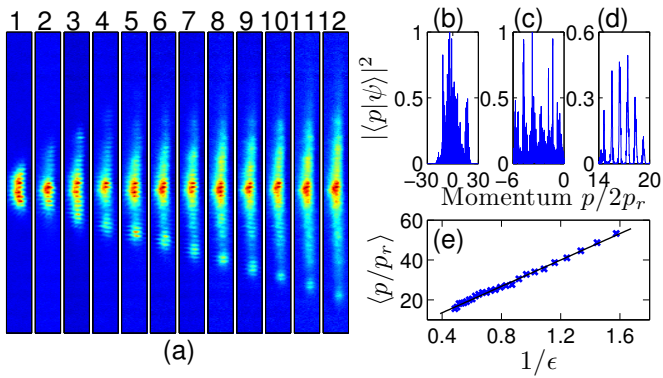}
\caption{(Color online) Behavior of transporting island. In (a), false color experimental images of momentum distribution are shown as a function of the number of kicks, with $\epsilon = 1$. Quantum simulations (with $\epsilon = 1$) illustrate the differences between the island and chaotic regions: the momentum wavefunction after 8 kicks is shown (arbitrary units) in (b), (c) and (d). Closeups of a section of the chaotic region (c) and the island region (d) are shown. In (e), experimental data of the mean momentum of island atoms is plotted as a function of $1/\epsilon$, where $\epsilon$ is the detuning from quantum resonance. All images are for phase modulation \{0,$\frac{2\pi}{3}$,0\}, $\beta = 0.43$ and $k = 3.2$.}
\label{fig3}
\end{figure}

Directed transport is observed in the case where the initial condition overlaps a transporting island. Fig.~\ref{fig3}(a) shows atoms ``on the island" being linearly accelerated in the downward direction with every kick. The island is narrow --- only 3--4 diffraction orders in width --- but contains up to 20\% of the atoms in the system. We measure an average acceleration of $3.8 \pm 0.2$ photon recoils per kick, in agreement with the classical prediction for a system of $j$ islands of:

\begin{equation}
\Delta J_{kick} = \epsilon\Delta\rho_{kick} = \frac{2\pi}{j}.
\label{eq:acceleration}
\end{equation}

\noindent For the three--island system of the \{0,$\frac{2\pi}{3}$,0\} phase modulation sequence, with $\epsilon = 1$, the classical prediction gives $\Delta p_{kick} = 4.2~p_r$. A quantum--mechanical simulation gives an equivalent result.

Our simulations reveal a striking difference between the chaotic and island regions. Fig.~\ref{fig3}(b) shows the simulated momentum wavefunction for our system. Zooming in on the chaotic region in (c) shows typical quantum chaotic behavior, with `jagged peaks' centered around integer momenta, offset by $\beta$. In contrast, zooming in on the island region in (d) shows near--perfect, narrow Gaussian peaks.

To first order, the island acceleration is independent of the kick--strength $k$, as indicated in (\ref{eq:acceleration}), provided that the phase--space does not bifurcate with changing $k|\epsilon|$. Fig.~\ref{fig3}(e) shows the average momentum of atoms in the island as a function of $1/\epsilon$. The strong linearity is in excellent agreement with the classical prediction of a $1/\epsilon$ dependence of the island momentum, from Equation~\ref{eq:acceleration}.

\begin{figure}
\includegraphics[width=250px]{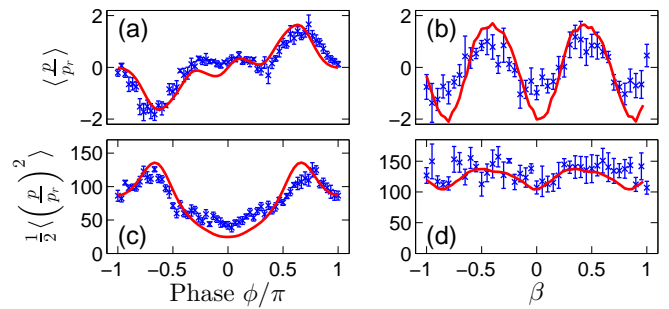}
\caption{(Color online) Experimental and theoretical net currents and energies as a function of phase modulation and of quasi--momentum $\beta$. Net momentum current (a) and average energy per atom (b) as a function of modulated phase $\phi$ for the phase sequence $\{0,\phi,0\}$, with $\beta = 0.43$. Net momentum current (c) and average energy per atom (d) as a function of quasi--momentum $\beta$, for the phase sequence $\{0,\frac{2\pi}{3},0\}$. Momentum is measured in single--photon recoil ($p_r$) units. Data collected after 8 kicks with $\epsilon = 1$. The kick-strength for (a) and (c) is calibrated to be $k = 3.0$; for (b) and (d), $k = 2.9$. The experimental data is overlaid with quantum simulations.}
\label{fig4}
\end{figure}

The reversible ratchet effect is clearly demonstrated in Fig.~\ref{fig4}(a). A phase scan is performed for the phase modulation $\{0,\phi,0\}$. A peak in the net momentum current of $\pm 1.8$ photon recoils after 8 kicks is observed, for $\phi \approx \pm\frac{2\pi}{3}$. The close match of our experimental data to the simulation, extending to the inflexion feature observed near zero--phase modulation, is indicative of the precise degree of control we have over the phase of the standing wave.

Figure~\ref{fig4}(c) demonstrates our ability to address individual islands in the pseudo--classical phase space. As the frequency difference between the beams is scanned, the quasi--momentum $\beta$ is altered and the initial condition is offset in $J$. Peaks are observed in the vicinity of $\beta = \pm 0.43$: this is the center of the transporting island as predicted pseudo--classically.

The energy plots of Figures~\ref{fig4}(b) and (d) further illustrate the transport enhancement brought about through the chosen phase modulation set. The transport peaks in the vicinity of $\beta = 0.43$, around the modulated phase $\phi \approx \frac{2\pi}{3}$. These figures further emphasize the agreement between the experimental results and the theoretical predictions.

\begin{figure}
\includegraphics[width=250px]{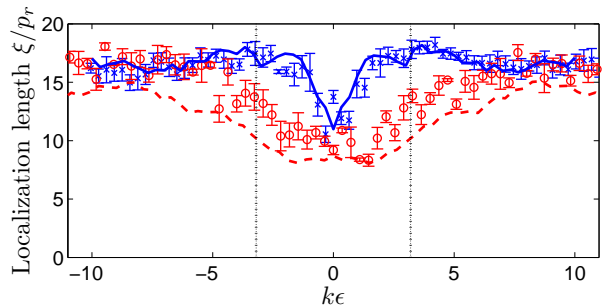}
\caption{(Color online) Experimental plot of the momentum--space localization length, measured in single--photon recoil units, as a function of $k\epsilon$. Blue crosses represent data from a phase modulation of $\{0,\frac{2\pi}{3},0\}$. Red circles represent data from an unmodulated kicked rotor. The data has been collected after 25 kicks around the $\ell = 2$ resonance, with $\beta = 0.2$. The kick--strength is calibrated to be $k = 5.7$ for the modulated experiments, and $k = 6.4$ for the unmodulated experiments. The dotted lines indicate $k\epsilon = \pm 3.2$. The results are overlaid with quantum simulations: solid blue indicates phase modulation, while dashed red indicates an unmodulated rotor. The simulations have been conducted with a high momentum cutoff of $\pm 45 p_r$ to mimic our finite experimental view-window.}
\label{fig5}
\end{figure}

The results shown in Fig.~\ref{fig5} indicate experimental observations of a significant peak in the localization length $\xi = \sqrt{\langle p^2\rangle/2}$ of chaotic atoms around $k|\epsilon| = 3.2$ due to the transporting islands generated by phase modulation. The data is collected after 25 kicks, well into the dynamical localization regime. For this experiment, we use $\ell = 2$, where the transporting island is centered at $\beta = 0$: a value of $\beta = 0.2$ means that the initial condition does not overlap the transporting island. This choice of quasi--momentum has the added advantage that the zero--phase modulation quantum resonance at 66.3 $\mu$s is negated~\cite{Currivan2009}. We record a localization length of the chaotic region enhanced by 40\%, when compared with the zero--phase--modulation equivalent.

We note that although the wavefunction clearly has two separate behavior regimes, as shown in Fig.~\ref{fig3}(c)--(d), the chaotic and island motions are by no means independent. Although the chaotic region is still subject to dynamical localization~\cite{Iomin2002}, the localization length is greatly increased by the phase modulation~\cite{Sadgrove2013}. This is well demonstrated in Fig.~\ref{fig5}, despite the initial condition not classically overlapping the transporting island, and the island classically not affecting the chaotic motion.

It is important to note the origin of the ratchet behavior that we have observed. We stress that there is no net biased force in our system: every kick, when considered in isolation, produces a symmetric momentum distribution about the initial momentum state. The asymmetric phase modulation that we impose induces interference effects that lead to the observed net momentum current. The concept of a well defined trajectory loses its meaning for quantum particles after the short (sub--kick period) Ehrenfest time~\cite{Izrailev1990}. Instead, interference from all sections of the phase space contributes to the final momentum distribution. This results in the chaotic region exhibiting greater transport and a longer localization length than in the unmodulated case. Furthermore, violation of a classical sum rule is possible~\cite{Schanz2001,Gong2004}, allowing for directed quantum transport even when the entire phase space is sampled~\cite{Gong2006}.

In summary, we have experimentally observed a transporting island structure originating from phase modulation of a pulsed optical standing wave, resulting in linearly accelerated atoms. The asymmetry of our chosen phase modulation set leads to unidirectional transporting islands in the $\epsilon$--classical phase space: a new type of quantum ratchet has been demonstrated. The net momentum current is tunable through the modulated phase and the quasi--momentum $\beta$. The phase modulation has been found to enhance the localization length of chaotic atoms, despite the chaotic region being classically independent of the island structure. All of our data is in excellent agreement with both $\epsilon$--classical theory and quantum simulations. This work could be extended by using a near non--interacting condensate, with our simulations indicating that a narrower initial momentum distribution results in an enhancement of a factor of five in the directed current.

The authors acknowledge the support of the University of Auckland Research Fund. We thank Sandro Wimberger for helpful discussions.

\bibliography{references}

\end{document}